\documentstyle[11pt,newpasp,twoside,psfig]{article}
\index{Anomalous X-ray Pulsar}
\index{4U~0142$+$61}

\def\edcomment#1{\iffalse\marginpar{\raggedright\sl#1\/}\else\relax\fi}
\reversemarginpar

\begin{document}
\title{{\it \bf Chandra}/HETGS Spectroscopy of the Anomalous X-ray Pulsar 
	4U~0142$+$61}
\author{Adrienne M. Juett, Herman L. Marshall, Deepto Chakrabarty, 
	Claude R. Canizares, and Norbert S. Schulz}
\affil{Center for Space Research and Department of Physics, 
	Massachusetts Institute of Technology, Cambridge, MA 02139}

\begin{abstract}
We report on a 25 ks observation of the anomalous X-ray pulsar 
4U~0142$+$61 with the High Energy Transmission Grating Spectrometer 
(HETGS) on the {\it Chandra X-ray Observatory}.  The continuum spectrum 
is consistent with previous measurements and is well fit by an absorbed 
power-law $+$ blackbody model with photon index $\Gamma=3.3\pm0.4$ and 
blackbody temperature $kT=0.418\pm0.013$~keV.  The pulsar frequency was 
0.1150966$\pm$0.0000017~Hz and the pulse fractions were between 8.7\% 
and 21\%, which are also consistent with past measurements.  No evidence 
was found for emission or absorption lines with an upper limit of 
$\approx$50~eV on the equivalent width of a broad feature in the 
2.5--13~\AA\/ (0.95--4.96~keV) range.  The absence of a proton cyclotron 
line strongly constrains magnetar atmosphere models and hence the magnetic 
field strength of the neutron star.  For the energy range given above, the 
allowed magnetic field strengths of 4U~0142$+$61 are 
$B<1.9\times 10^{14}$~G and $B>9.8\times 10^{14}$~G.  
\end{abstract}

\section{Introduction}
Anomalous X-ray pulsars (AXPs) were suggested as a class by Mereghetti 
\& Stella (1995) based on their observational properties.  AXPs have a 
tight range of spin periods, 6--12 s, luminosities of order 
10$^{34}$--10$^{35}$ erg s$^{-1}$, and a soft X-ray spectrum described 
by a blackbody of temperature 0.2--0.6~keV and a power-law of index 2--4 
(see, Gavriil \& Kaspi 2001, and references therein).  These sources also
 undergo relatively steady spin-down, have faint or unidentified optical 
counterparts, and have no evidence of orbital motion.  The properties of 
AXPs do not point to a single explanation of the systems, but rather 
several different models have been suggested to account for the 
observational properties.  These models fall into two general categories: 
(1) accretion models where the material is from either a very low-mass 
companion or from a fallback disk (see, e.g., van Paradijs, Taam, \& van 
den Heuvel 1995), and (2) magnetar models which suggest that AXPs are 
ultra-magnetized (B$\approx$10$^{14}$--$10^{15}$), isolated neutron stars 
(see, e.g., Thompson \& Duncan 1996).  Recently, atmospheric modeling of 
magnetars predict that a broad ($\Delta E/E\approx1$) proton cyclotron 
absorption line should be apparent in the X-ray spectrum of magnetars 
(Ho \& Lai 2001; Zane et al. 2001).  High-resolution X-ray 
spectroscopy should be able to identify the proton cyclotron feature 
or at least place limits on the magnetic field strengths of the AXPs.

4U~0142$+$61 is a member of this class, with spin period of 8.7~s 
(e.g., Gavriil \& Kaspi 2001).  X-ray spectra obtained with {\it ASCA\/} 
and {\it BeppoSAX\/} were well fitted by a two component model of a 
0.4 keV blackbody and a $\Gamma\approx$3.7 power-law 
(White et al. 1996; Israel et al. 1999; Paul et al. 2000).  Recently, 
Hulleman, van Kerkwijk, \& Kulkarni (2000) identified an optical 
counterpart based on the {\it Einstein\/} position.  

\section{Chandra/HETGS Results}
We observed 4U~0142$+$61 with {\it Chandra\/} on 2001 May 23 for 25 ks 
using the High Energy Transmission Grating Spectrometer (HETGS; 
Canizares et al. 2001, in preparation) and the Advanced CCD Imaging 
Spectrometer (ACIS; Garmire et al. 2001, in preparation). Traditionally, 
the source has been fit with a power-law and a blackbody both absorbed
by neutral gas in the interstellar medium.  The results of the 
{\it Chandra\/} fit, shown in Figure 1, are consistent with previous 
observations of 4U~0142$+$61.  The best fit parameter values are: 
photon index $\Gamma=3.3\pm0.4$, power-law normalization at 1 keV 
A$_{1}=0.10\pm0.05$ photons keV$^{-1}$ cm$^{-2}$ s$^{-1}$, blackbody 
temperature $kT=0.418\pm0.013$~keV and blackbody radius 
$R_{\rm bb}=(2.0\pm0.2) d_{\rm kpc}$ km. 

\begin{figure}[t]
\centerline{\psfig{file=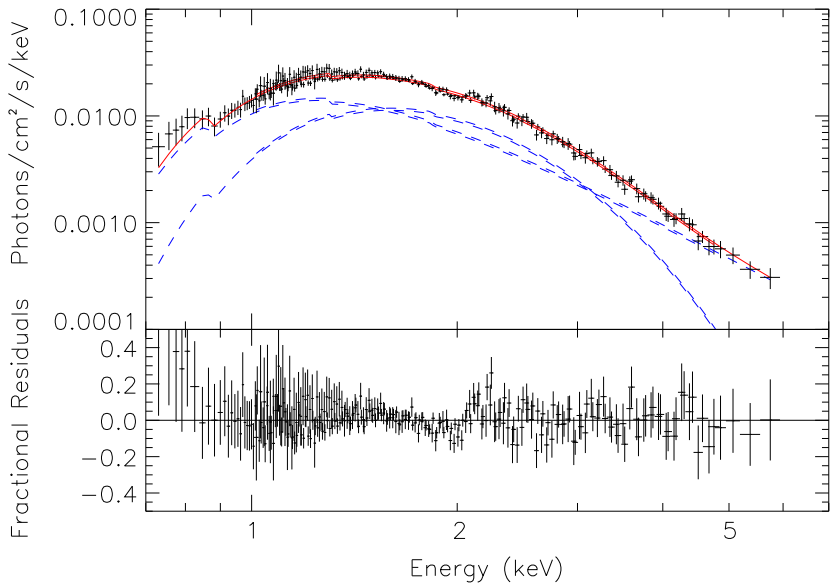,width=0.9\linewidth}}
\caption{(upper panel) Unfolded energy spectrum of 4U~0142$+$61 fit with a power-law and a blackbody model.  The contributions of the power-law and blackbody components are also shown.  (lower panel) Fractional residuals ([data-model]/model) from the power-law and blackbody fit.}
\label{fig:1} 
\end{figure}

With a well determined continuum model, we then looked for absorption 
and emission features in the high-resolution spectrum.  We fit Gaussian 
models to the fractional residuals ([data-model]/model) from the continuum 
fit.  The central energies and widths of the Gaussian components were 
fixed, while the normalizations were fitted.  To look for features that 
had been predicted in the magnetar models (Ho \& Lai 2001; 
Zane et al. 2001), the sigma of the Gaussian was chosen to vary with 
energy ($\sigma =0.1\times E$).  The Gaussian model was fit to the data 
centering at every wavelength point.  We were able to determine the best 
fit amplitude and standard deviation of this result as well as the 
significance of each feature.  There were no features with a significance 
greater than 4$\sigma$. Magnetar models predict equivalent width values 
of 0.70--0.75$E$ which are significantly above our upper limits of 
$\approx$50~eV.  Thus, there are no features that could be attributed 
to a proton cyclotron feature in the range 2.5--13~\AA\/ 
(0.95--4.96~keV).  For a more detailed account of our analysis, see 
Juett et al. (2001).

Our timing analysis used the dispersed events identified at the first 
and second orders.  The event arrival times were barycentered and 
randomized within the frame time of 1.84 s.  A lightcurve was created 
with 2 s bins, from which a power spectrum was made.  The power 
spectrum showed both the fundamental and the second harmonic of the 
8.7~s period.  The best fit frequency of 0.1150966$\pm$0.0000017 Hz 
was measured from the second harmonic and is consistent with the 
ephemeris of Gavriil \& Kaspi (2001) derived from long term monitoring 
with {\it RXTE}.  To look at the pulse profile as a function of energy, 
the events were filtered into five energy bands using the energy column 
of the event file.  The lightcurves for each energy band were then 
folded on the best fit period and the pulse fraction was determined, 
defined as PF=$(F_{max}-F_{min})/(F_{max}+F_{min})$ (see Fig.~2).  An 
energy-independent pulsed fraction is excluded at the 1.7$\sigma$ level. 

\begin{figure}[t]
\centerline{\psfig{file=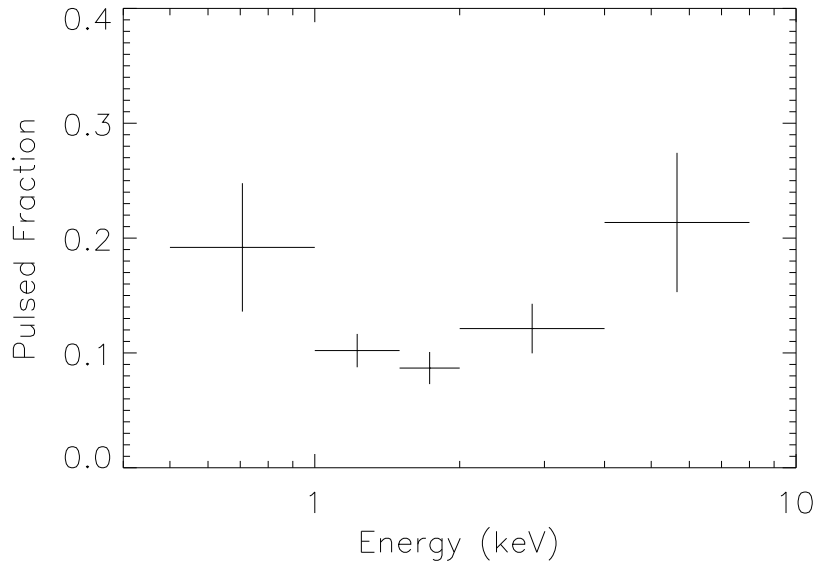,width=0.9\linewidth}}
\caption{Pulsed fraction as a function of energy, defined as 
$(F_{max}-F_{min})/(F_{max}+F_{min})$, where $F_{max}$ and $F_{min}$ 
are the maximum and minimum values of the observed photon flux.  
An energy-independent pulsed fraction is excluded at the 1.7$\sigma$ level.}
\label{fig:2}
\end{figure}

\section{Discussion}
The {\it Chandra\/} spectrum of 4U~0142$+$61 is well fit by a two 
component power-law and blackbody model absorbed by the interstellar 
medium, which is consistent with previous observations of the source 
(White et al. 1996; Israel et al. 1999; Paul et al. 2000).  The position 
we obtained from the {\it Chandra}/HETGS 0th order confirms the optical 
identification.  No significant features were found in a search of the 
high-resolution spectrum.  We can place a very conservative limit on the 
equivalent width of any broad feature of $\approx$50 eV in the range 
2.5--13~\AA\/ (0.95--4.96~keV).  The limits on the equivalent width of 
any absorption features place strong constraints on the neutron star 
atmosphere models.  Atmosphere models of highly magnetized neutron 
stars indicate that proton cyclotron absorption would produce a broad 
feature at an energy $E_{B}=0.63y_{g}(B/10^{14}\:{\rm G})$~keV, where 
$y_{g}=(1-2GM/c^{2}R)^{1/2}$ is the gravitational redshift factor 
(Ho \& Lai 2001; Zane et al. 2001).  For the range we can reasonably study, 
0.95--4.96~keV, our equivalent width limits are much lower than the 
predicted equivalent widths of order 0.70--0.75$E_{B}$.  If we assume 
that the gravitational redshift to the surface is 0.2, then the range 
of disallowed magnetic field strengths is (1.9--9.8)$\times 10^{14}$ G.

\end{document}